\documentstyle[12pt,a4,epsfig]{article}
\newcommand{\rcite}[1]{{\cite{#1}}}

\newcommand{\tref}[1]{{\ref{#1}}}
\newcommand{\rlabel}[1]{{\label{#1}}}
\newcommand{\rbibitem}[1]{\bibitem{#1}}
\newcommand{\be}{\begin{equation}}
\newcommand{\ee}{\end{equation}}
\newcommand{\ba}{\begin{eqnarray}}
\newcommand{\ea}{\end{eqnarray}}
\begin{document}
\begin{titlepage}
\begin{flushright}
LU TP 97/06\\
hep-ph/9704233\\
April 1997
\end{flushright}
\vfill
\begin{center}
{\large\bf Chiral Perturbation Theory and Threshold Corrections}\footnote{
Invited
talk given at the ``MAX-lab Workshop on the Nuclear Physics Programme
with Real Photons below 200~MeV,'' Lund, March 10-12, 1997}\\[2cm]
{\bf Johan Bijnens}\footnote{email: \tt bijnens@thep.lu.se}\\[0.5cm]
Department of Theoretical Physics, University of Lund,\\
S\"olvegatan 14A,
S 223 62 Lund, Sweden
\end{center}
\vfill
\begin{abstract}
I give a short overview of Chiral Perturbation Theory, its underlying
assumptions and underpinnings. A few examples are included. 
\end{abstract}
\vfill
\end{titlepage}
\section{Introduction}
In this talk I will give a short introduction to Chiral symmetry and the
Goldstone theorem. These we then combine into Chiral Perturbation Theory(CHPT).
This is a systematic tool to solve the Ward identities of Chiral symmetry. Its
main advantage over the use of models is that it is a theory, i.e. higher order
corrections can in principle be calculated and the convergence of the
expansion tested. It is also systematic, there are no {\em hidden} assumptions.
$\pi\pi$ scattering will be used as an example here.

We will then go beyond PCAC and show more examples,
$\gamma\gamma\to\pi^+\pi^-$ where a naive estimate would have worked
and $\gamma\gamma\to \pi^0\pi^0$ where the naive expectation did not work.

The last section is an
extremely short review of pion photoproduction. This is mainly the
work of V.~Bernard et al..

\section{Chiral Symmetry}

\subsection{Definition and Goldstone Theorem}
\rlabel{goldstoenthm}

In Quantum Chromodynamics (QCD) we have quarks, if they have the same
(or nearly the same) mass we have a symmetry by interchanging them. For
the case of two flavours this is known as isospin. It is a continuous symmetry,
$SU(2)_V$. A generalization to three flavours is the Gell-Mann-Okubo eightfold
way where the symmetry is enlarged to $SU(3)$.

In the case of massless particles this symmetry is larger. The underlying
reason is that left and right handed particles are really distinct entities.
As shown in Fig. \tref{figchiral} by overtaking a massive particle you change
the direction of its momentum but not of its angular momentum. So you flip its
helicity. Since you cannot overtake a massless particle the two helicities
are fully separate.
\begin{figure}
\begin{center}
\setlength{\unitlength}{1mm}
\thicklines
\begin{picture}(120,50)
\put(0,39){$m\ne 0$~:}
\put(20,40){\vector(1,0){30}}
\put(51,39){$p$}
\put(35,40){\oval(10,10)[l]}
\put(35,35){\vector(1,0){1}}
\put(30,30){right}
\put(48,42){\parbox[b]{4cm}{\begin{center}Lorentz\\Transformation
		  \end{center}}}
\put(65,39){$\Longrightarrow$}
\put(10,0){
\put(100,40){\vector(-1,0){30}}
\put(85,40){\oval(10,10)[r]}
\put(85,35){\vector(-1,0){1}}
\put(101,39){$p'$}
\put(80,30){left}}
\put(0,-30){
\put(0,39){$m = 0$~:}
\put(20,40){\vector(1,0){30}}
\put(51,39){$p$}
\put(35,40){\oval(10,10)[l]}
\put(35,35){\vector(1,0){1}}
\put(30,30){right}
\put(48,39){\parbox{4cm}{\begin{center}You cannot\\go faster\\than light
		  \end{center}}}
%\put(65,39){$\Longrightarrow$}
\put(10,0){
\put(77,40){\vector(1,0){16}}
\put(85,40){\oval(10,10)[l]}
\put(85,35){\vector(1,0){1}}
\put(101,39){$p'$}
\put(80,30){right}} }
\end{picture}
\end{center}
\caption[]{Lorentztransformations couple left and right helicities for
massive particles by changing the sign of he momentum. This is not
possible for massless particles.}
\rlabel{figchiral}
\end{figure}
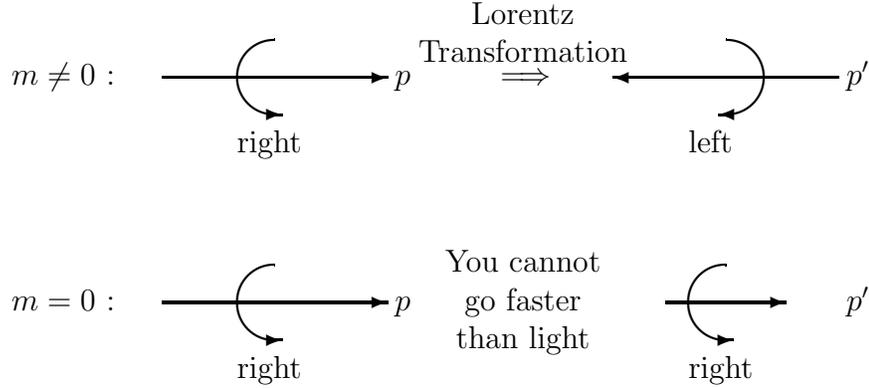 
As a consequence isospin gets doubled to the chiral symmetry group
$SU(2)_L\times SU(2)_R$ with a separate ``isospin'' for both helicities.
In the case of 3 massless flavours the symmetry group is
$SU(3)_L\times SU(3)_R$.

This symmetry group is however not manifest in nature at all. E.g., the
parity partners vector mesons
$a_1$ and $\rho$ have masses of 1230~MeV and 770~MeV respectively,
the proton-neutron and their partners, the $S_{11}$, have masses of about
940~MeV versus 1535~MeV. So the question is: where is the symmetry ?

The Wigner-Eckart theorem we all learned in our quantum mechanics course has a
loophole, the Goldstone theorem uses this loophole. The usual proof of the
Wigner-Eckart theorem assumes that the vacuum, or lowest energy state,
is unique. Once we drop this requirement, the Wigner-Eckart theorem
is no longer valid. Let me show it in a somewhat schematic way. For
a symmetry generated by $Q$ we have two related states,
$\alpha$ created by the creation operator $a^\dagger$ and
$\beta$ created by $b^\dagger = e^{i\gamma Q}a^\dagger e^{-i\gamma Q}$.
$Q$ and $e^{i\gamma Q}$ commute with the Hamiltonian $H$ since it generates
a symmetry. Then:
\ba
m_\beta &=& \langle 0 | b H b^\dagger | o\rangle\nonumber\\
&=&\langle 0| e^{i\gamma Q}a e^{-i\gamma Q}H e^{i\gamma Q} a^\dagger
e^{-i\gamma Q}
|0\rangle\nonumber\\
&=&\langle0|e^{i\gamma Q}a H a^\dagger e^{-i\gamma Q}|0\rangle\nonumber\\
&&\hspace{-1.5cm}\mbox{if }Q|0\rangle=|0\rangle
\mbox{, the vacuum is a singlet under the symmetry}\nonumber\\
&=&\langle0|a H a^\dagger |0\rangle\nonumber\\
&=&m_\alpha
\ea
So if there are several vacua we can have a symmetry group and
$m_\alpha\ne m_\beta$.

The underlying symmetry still has lots of consequences.
See \rcite{goldstone}. A naive proof goes as follows:\\
If you have a continuous symmetry generated by a generator $Q$. The effect of
the other vacua that have to be chosen at each point in space time can be
described by a field $\phi(x)$. The vacuum in a point $x$ is given by
$\displaystyle e^{i\phi(x)Q}|0\rangle$ where $|0\rangle$ is a reference
vacuum. The symmetry is still a global symmetry, so a rotation that is the
same in all space time cannot do anything. So
$\displaystyle e^{i\phi(x)Q}|0\rangle$ and
$\displaystyle e^{i(\phi(x)+\alpha)Q}|0\rangle$ describe the same state.
The Lagrangian (or more precisely the action) should be the same for
$\phi(x)$ and $\phi(x)+\alpha$ with $\alpha$ a constant. The dependence
on the field $\phi$ can thus only happen via derivatives or $\phi$ can
only occur as $\partial_\mu\phi$. Thus mass terms are excluded and interactions
vanish at zero energy and momentum. The massless particle described by $\phi$
is called a Goldstone boson.

\subsection{Chiral Perturbation Theory}

For CHPT we use the Goldstone theorem for the chiral symmetry. The
symmetry which is broken is the axial part of the chiral symmetry. The diagonal
vector subgroup remains unbroken. For $N_f$ flavours this means that there
are $N_f^2-1$ broken symmetries or we will get $N_f^2-1$ Goldstone
bosons. The interactions are weak at low energies. We can therefore do
a systematic expansion in the number of derivatives and have a well defined,
consistent perturbation theory.
In the case of two flavours we can define a four-vector $\vec{U}$ containing
the 3 Goldstone bosons $\pi^i$ and a lowest order Lagrangian:
\ba
\vec{U}&=&\left(\sqrt{1-\frac{\vec{\pi}^2}{F^2}},\frac{\pi^1}{F}
,\frac{\pi^2}{F},\frac{\pi^3}{F}\right)
\nonumber\\
\vec{\chi}&=&2 B\left(s^0,p^i\right)
\nonumber\\
{\cal L}_2 &=&\frac{F^2}{4}\left(\nabla_\mu\vec{U}\cdot\nabla^\mu\vec{U}
+\vec{\chi}\cdot\vec{U}\right)\,.
\rlabel{L2}
\ea
The covariant derivative $\nabla_\mu$ is defined in \rcite{GL1}.
The field $\chi$ contains the quark masses. The extension to 3 flavours is
in \rcite{GL2}. This Lagrangian contains at tree level
a very large part of all the PCAC predictions of the sixties. The reason
for using external fields is explained in \rcite{GL1,GL2}. For
a simple example explaining the advantage see \rcite{korea}.

\subsection{Powercounting and $\pi\pi$-scattering.}

As argued in the previous subsection we have a well defined expansion
in terms of derivatives (and quark masses). This was proven in a simple
way in Weinberg's paper\rcite{weinberg}. Let us show the arguments at the
example of $\pi\pi$ scattering. The classes of diagrams are shown in
Fig. \tref{figpipidiag}.
\begin{figure}
\begin{center}
\leavevmode\epsfig{file=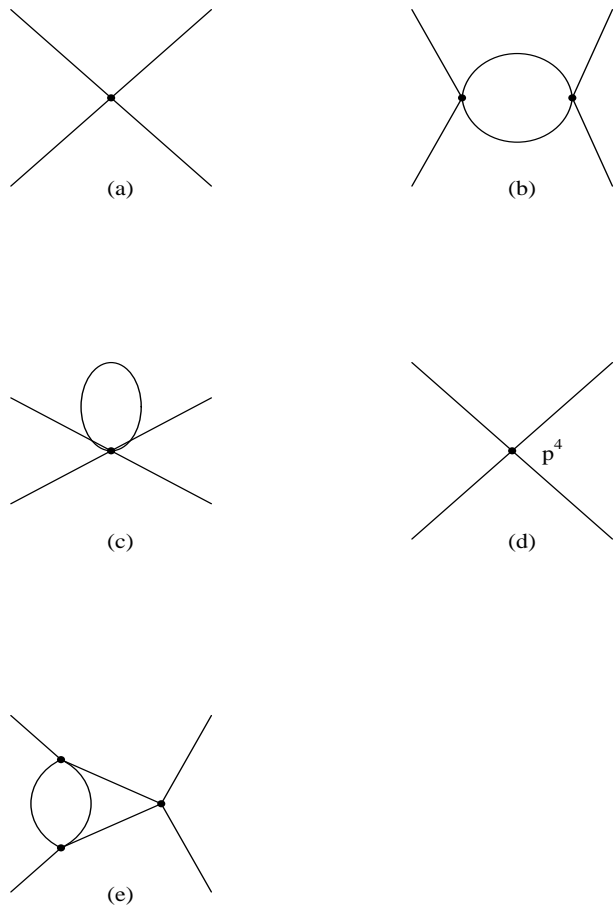,height=12cm,width=8cm}
\end{center}
\caption{Some diagrams contributing to $\pi\pi$ scattering.}
\rlabel{figpipidiag}
\end{figure}
A vertex contains two derivatives, so two
powers of momenta, $p^2$, a propagator is the inverse of the kinetic term,
it has dimension $-$2, or $p^{-2}$ and a loop integral, $d^4p$, has
dimension 4 or $p^4$. The lowest order term of Fig. \tref{figpipidiag}a
is order $p^2$. The loop diagrams of (b) and (c) are $p^4$ and the
two-loop diagram of (e) is $p^6$. The convergence of this expansion
works quite well. The lowest order\rcite{weinberg1}, the $p^4$\rcite{GL1}
and the $p^6$ calculation\rcite{pipi2loop}
 converge up to about 500~MeV or so. The
combination measurable in $K_{l4}$ decays is shown in Fig. \tref{figpipiphase}.
The data are from \rcite{rosselet}.
\begin{figure}
\begin{center}
\leavevmode\epsfig{width=12cm,file=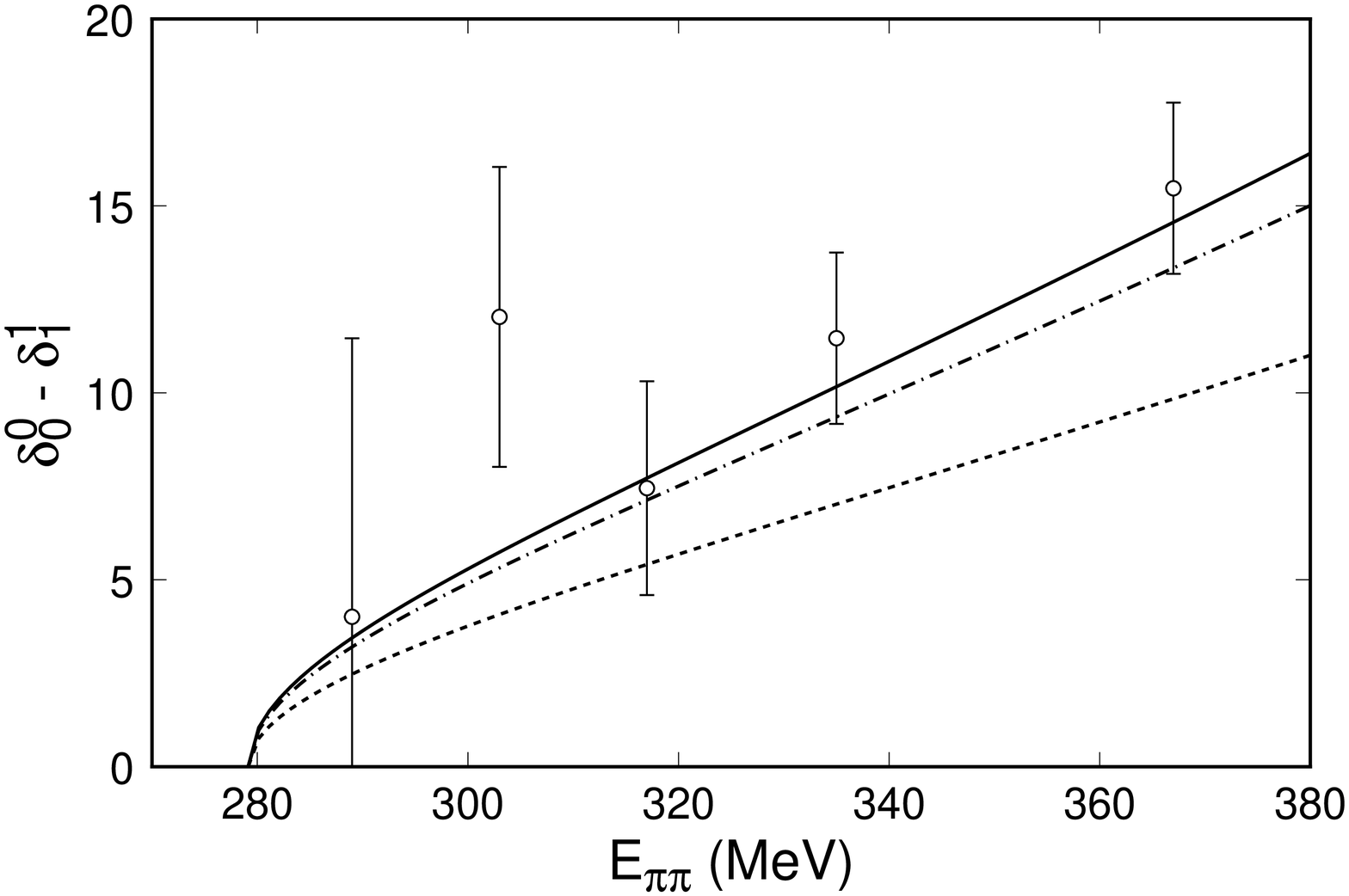}
\end{center}
\caption{The $\pi\pi$-phase shift measurable in $K_{l4}$ together with the
CHPT results\protect{\rcite{pipi2loop}}.}
\rlabel{figpipiphase}
\end{figure}
We can also look at one of the scattering lengths at threshold to see the
convergence:
\ba
a^0_0 &=& 0.156 (\mbox{tree}) +
\overbrace{0.039 (L) + 0.005(\mbox{anal})}^{\mbox{1-loop}}
\nonumber\\&&
+\overbrace{0.013(L^2+l_i L)+0.003(L)+0.001(\mbox{anal})}^{\mbox{2-loop}}\,.
\ea
Here the first term is the tree level. The symbol $L$ stands for the
nonanalytic contribution proportional to $L=\log(m_\pi^2/\mu^2)$,
$L^2$ for those proportional to the square and $l_i L$ for the one-loop
contributions with a $p^4$-vertex in the diagram.
The unknown $p^6$ coefficients only contribute to the last term.
In this particular case the nonanalytic pieces dominate so the uncertainty
due to higher order terms is rather small.

\section{Examples}

\subsection{$\gamma\gamma\to\pi^+\pi^-$}

This is a process where everything works as we expect. The Born term
is just scalar electro dynamics and dominates the cross section within the
whole domain of validity of CHPT. The $p^4$ part\rcite{BC} gives
a reasonable correction and the $p^6$ calculation
\rcite{burgi} has even smaller effects.
See the figures in \rcite{burgi}.

\subsection{$\gamma\gamma\to\pi^0\pi^0$}

If we would have been naive we would have said: There are no terms in the
action contributing to order $p^2$ and $p^4$ (this was known as the
Veltman-Sutherland theorem) but there are terms of order $p^6$. An
order of magnitude estimate of their coefficient would lead to a cross
section of order 0.04~nb. However the nonanalytic pieces are nonzero
at order $p^4$\rcite{BC,DHL} and give a cross section of a few nanobarn.
This agreed roughly with the measurements. The $p^6$ corrections were later
calculated and still found to be significant\rcite{BGS}. See
Fig. \tref{figggpp}.
\begin{figure}
\begin{center}
\leavevmode\epsfig{file=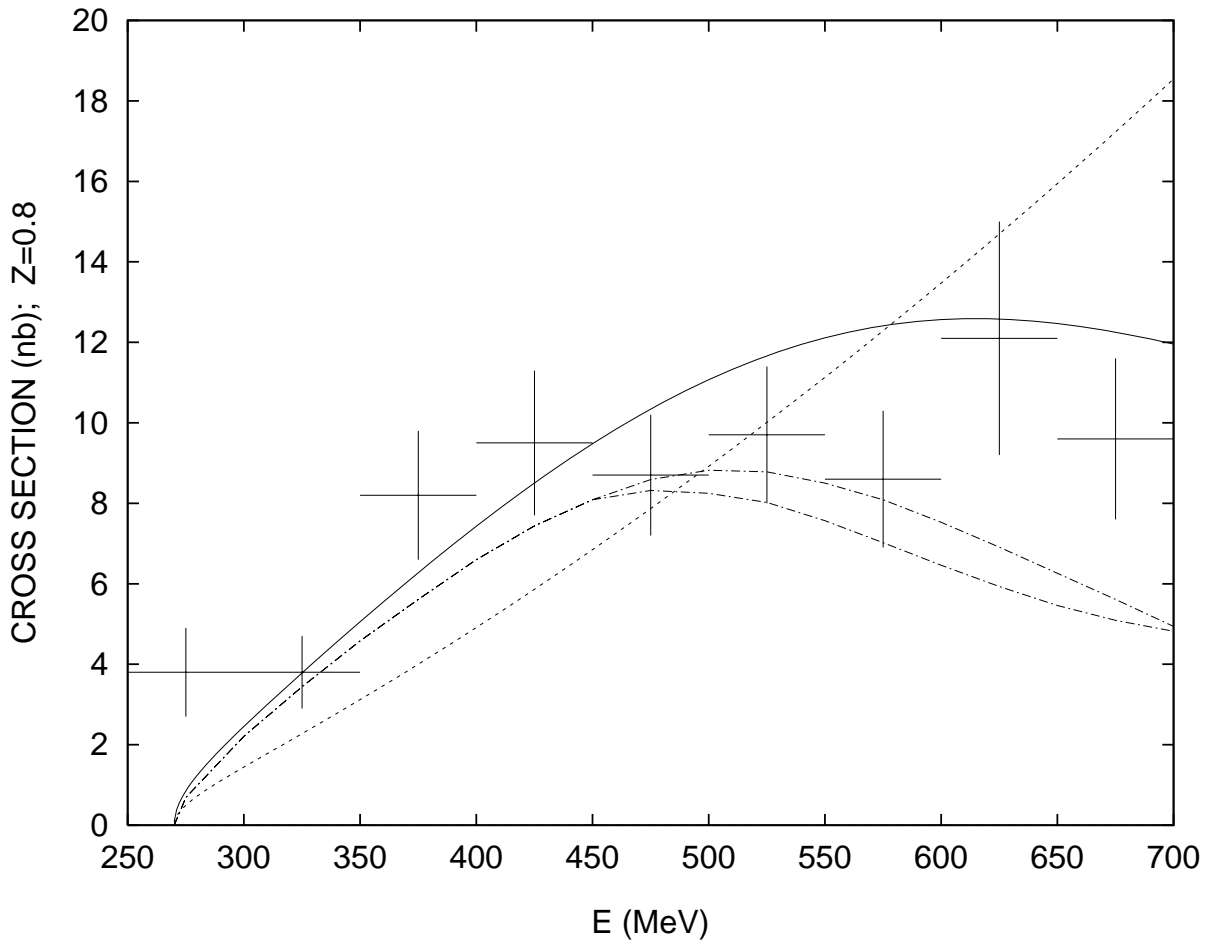,width=12cm,angle=-90}
\end{center}
\caption[]{The cross section $\gamma\gamma\to\pi^0\pi^0$. Dashed line is $p^4$,
full line $p^6$ and the crosses the data\protect{\rcite{CB}}. Figure taken from
\protect{\rcite{BGS}}.}
\rlabel{figggpp}
\end{figure}

So, the {\em Low Energy Theorem} was wrong, the amplitude is nonanalytic
and the loops gave the correct order of magnitude. So why were there still
very large $p^6$ corrections ?

The reason is that the underlying isospin amplitudes, I=0,2, are both very
large. For the charged pion production they add up so we only find moderate
corrections to the total. For the neutral pion production they cancel,
but they have different higher order corrections, I=0 has large
rescattering effects, I=2 has not. Differences of large numbers tend
to be much more sensitive to higher orders.

\section{Pion Photoproduction}

Similar effects as described in the previous section happen here.
For $E_{0+}$ in $\gamma p\to\pi^0 p$ there was a similar wrong
low energy theorem\rcite{bgetal}. All higher order corrections up to $p^4$
have by now been calculated by the group of Bernard, Kaiser and Mei\ss{}ner.
Examples of badly converging, like $E_{0+}$ in neutral,
and well converging quantities, like most quantities in charged
pion photoproduction, can
all be found. In the talk I presented the examples but much more detailed
discussions can be found in \rcite{bkmreview}.

\section{Conclusions}

Chiral perturbation provides a well defined framework for low energy
hadronic physics. It cleanly separates model aspects, which here are
estimates of the constants, from the basic aspects
of chiral symmetry. It is very successful in describing the data in its regimes
of applicability. However it should be kept in mind that it is
a perturbative low energy expansion. At higher energies we can still
use the principles of CHPT for organizing our thoughts but we should
beware of numerical results.

\section*{Acknowledgments}
I would like to thank the organizers for a pleasant meeting.

\end{document}